\documentclass[12pt]{iopart}
\usepackage{graphicx}
\usepackage{epstopdf}
\usepackage{color}
\usepackage{amsthm, amssymb, latexsym}
\usepackage{cite}

\renewcommand{\figurename}{Figure}

\begin{document}

\title[]{Amplitude modulation and surface wave generation in a complex plasma monolayer}

\author{Srimanta Maity$^{1*}$ and Garima Arora$^{2+}$}

\address{$^{1}$ELI Beamlines Facility, The Extreme Light Infrastructure ERIC, 
Za Radnicí 835, 25241 Dolní Břežany, Czech Republic \\}
\address{$^{2}$Department of Pulse Plasma Systems, Institute of Plasma Physics of the Czech Academy of Sciences, U Slovanky 2525/1a, 18200 Prague, Czech Republic }
\ead{$^{*}$srimantamaity96@gmail.com, $^+$garimagarora@gmail.com}
\vspace{10pt}
\begin{abstract}

The response of a two-dimensional plasma crystal to an externally imposed initial perturbation has been explored using molecular dynamics (MD) simulations. A two-dimensional (2D) monolayer of micron-sized charged particles (dust) is formed in the plasma environment under certain conditions. The particles interacting via Yukawa pair potential are confined in the vertical ($\hat z$) direction by an external parabolic confinement potential, which mimics the combined effect of gravity and the sheath electric field typically present in laboratory dusty plasma experiments. An external perturbation is introduced in the medium by displacing a small central region of particles in the vertical direction. The displaced particles start to oscillate in the vertical direction, and their dynamics get modulated through a parametric decay process. Consequently, beats generate in the vertical motion of the particles. It has also been shown that the same motion is excited in the dynamics of unperturbed particles as they are coupled via pair interactions. A simple theoretical model is provided to understand the origin of the beat motions of particles. Additionally, in our simulations, concentric circular wavefronts propagating radially outward are observed on the surface of the monolayer. The physical mechanism and parametric dependence of the observed phenomena are discussed in detail. It has been shown that the generated surface wave follows the dispersion relation of a transverse shear wave. This research provides insight into complex plasma crystals from the perspective of soft matter. 

\end{abstract}

\section{Introduction}\label{intro}

Complex plasma has been proven to be a remarkable medium to study waves \cite{merlino201425,morfill2003focus}, crystallization \cite{thomas1994plasma,chu1994direct,hayashi1994observation}, phase transition\cite{schweigert1998plasma,maity2019molecular,maity2022parametric}, cluster formation\cite{melzer2010finite,maity2020dynamical,deshwal2022chaotic}, crystal cracking\cite{maity2018interplay}, lane formation \cite{killer2016phase}, and many more. A low-power laser and standard camera are sufficient to visualize,  perturb and capture the dynamics of the medium due to its response with unique length (of the order of millimeters) and time scales (of the order of milliseconds). Complex plasma is a system of micron or submicron-sized dust particles suspended in the plasma environment. The suspended dust particles acquire large amounts of charge, which leads to much higher Coulomb potential energy with respect to the average thermal energy of particles. The medium exhibits from a fluid to a crystalline phase depending upon the ratio between Coulomb potential energy and dust thermal energy. Complex plasma is also considered a new state of soft matter where the medium's equilibrium properties and dynamical response depend upon the external conditions \cite{morfill2009complex}. 

Complex plasma medium has been shown to sustain various waves, e.g.,  linear waves such as dust acoustic waves (DAW)\cite{merlino201425}, transverse shear waves\cite{kaw1998low}, and nonlinear waves such as solitons \cite{kumar2017observation,arora2019effect,arora2021experimental} and shocks \cite{arora2020excitation}. Various nonlinear structures \cite{shukla2003solitons}, e.g., voids \cite{bailung2018characteristics}, Mach cones \cite{samsonov1999mach}, vortex \cite{bailung2020vortex}, etc., have also been reported in a dusty plasma medium. The first DAW was theoretically predicted by Shukla \textit{et. al.} \cite{rao1990dust} and experimentally realized by D'Angelo and Merlino \cite{merlino201425} in liquid and gaseous states. Kaw and Sen \cite{kaw1998low} predicted the new mode called transverse shear wave using a generalized hydrodynamic (GHD) model in dusty plasma when the medium is in a strongly coupled fluid state. Pramanik \textit{et. al.} \cite{pramanik2002experimental} confirmed this transverse shear wave experimentally in three-dimensional strongly coupled dusty plasma fluid.

The research on dust lattice waves has also been running alongside \cite{melandso1996lattice} since the discovery of plasma crystals. 
  Typically, dust lattice waves are excited in the laboratory by applying a modulating voltage to the wire near the dust crystal \cite{pieper1996dispersion}. Another novel technique introduced to excite lattice waves is using the radiation pressure of a laser which does not perturb the plasma environment and is considered far better than the modulated voltage technique. Nunomoura \textit{et. al.} \cite{nunomura2000transverse} used the same method to excite the in-plane transverse shear wave in monolayer dusty plasma crystal. The self-excited out-of-plane dust oscillations leading to instability were observed in Ref. \cite{nunomura1999instability}. In short, two types of waves were so far shown in the 2D monolayer plasma crystal, i.e., longitudinal waves called dust lattice waves, where the particle's motion is in the same direction of wave propagation, and in-plane transverse shear waves, in which particles oscillate perpendicular to the wave propagation. However, the out-of-plane transverse shear wave has not been observed in the past. In the present work, we have demonstrated for the first time the generation of an out-of-plane wave in a dusty plasma monolayer which follows the dispersion of a transverse shear wave. Amplitude modulation of the initial external perturbation is shown to be responsible for the generation of this wave.
  
 Amplitude modulation of dust lattice waves (DLW) was numerically studied by Melandso \cite{melandso1996lattice}. A theoretical model for the slow modulation of the DLW was reported by Amin\textit{et. al.} \cite{amin1998amplitude} by introducing a fast and slow motion of dust plasma as an initial condition. In the present work, we have observed a self-excited amplitude modulation of an initial external perturbation. The amplitude modulation generates surface waves, in which particles of the monolayer exhibit out-of-plane oscillatory motions. In our study, the external perturbation is introduced by displacing the small region of particles at the center of the monolayer in the downward ($-\hat z$) direction. This method can be easily realized in the laboratory, e.g., by focusing a laser pulse perpendicular to the plane of a complex plasma monolayer. Recently, in our previous work \cite{maity2022parametric}, we used the same technique and showed a first-order phase transition in the 2D dust crystal induced by parametric decay instability above a threshold value of initial displacement. In the present work, we introduce a small initial displacement to the particles so that the medium remains in the crystalline state. It has been found that the amplitude of initially displaced particles modulates via a parametric decay process and generates beats motion. A similar motion is also observed in the dynamics of unperturbed particles. The parametric decay process has been observed in many other aspects of plasma physics, inertial \cite{weber2015temperature} and magnetic confinement fusion \cite{oosako2009parametric}, and laser-plasma interactions \cite{maity2022mode}. In our simulations, the beat motion of particles creates a collective effect generating circular transverse wavefronts called surface waves, which propagate radially outwards from the center of the monolayer. The wave's characteristics have also been studied by changing the confinement potential as well as the dust density of the 2D monolayer. The atomistic picture of the wave reveals a mixture of longitudinal and transverse motion of particles similar to the surface wave in different media. The dispersion analysis of the surface wave has also been carried out and found that it follows the same dispersion characteristics as the theoretically predicted transverse shear wave. A simple theoretical model is also provided in support of our simulation results.
            
The whole paper is divided into different sections. In Sec. \ref{simu}, MD simulation details are described. In Sec. \ref{ampd}, the amplitude modulation process on the perturbed region in response to the external perturbation is discussed. The dependence of amplitude modulation phenomena on various system parameters, e.g., confinement frequency, dust density, and radius of the externally perturbed region, is described in various subsections. Section \ref{model} provides a theoretical model supporting our simulation observations. The collective response of the medium to the external perturbation is described in Sec. \ref{swg}. At last, all the results from simulation and modeling are summarised in Sec. \ref{summary}.   

\section{Simulation Details}\label{simu}

In this work, three-dimensional (3D) molecular dynamics (MD) simulations have been performed using an open-source massively parallel classical MD code LAMMPS \cite{plimpton1995fast}. Initially, ten thousand charged point particles (dust grains) interacting with each other via Yukawa pair potential have been distributed randomly inside a rectangular simulation box. In our simulation, we have considered periodic boundary conditions in all three directions. In our study, we have considered the charge ($Q$) and mass ($m_d$) of the dust grains to be $Q = -1000e$ and $m_d = 1.0\times 10^{-13}$ kg, respectively. Here, $e$ represents the magnitude of an electronic charge. In the vertical ($\hat z$) direction, a parabolic electrostatic potential $V_{ext} = (m_d\omega_v^2/2Q)(z-L_z/2)^2$ has been applied, which provides the vertical confinement of charged dust particles. Here, $\omega_v$ and $L_z$ represent parabolic confinement frequency and length of the simulation box along $\hat z$, respectively. The equation of motion of any $i$th particle can be expressed as,

\begin{equation}
 m_d\frac{d^2\mathbf{r}_i}{dt^2}= -Q\sum_{j=1}^{N-1}\nabla U(r_{i,j}) - Q\nabla V_{ext}  
\end{equation}
, where $r_i$ and $r_j$ are the positions of the $i$th and $j$th
particles at a particular time, respectively, and $U(r_{i,j})=(Q/4\pi\epsilon_0|\bf{r}_j-\mathbf{r}_i|)\exp{(-|\mathbf{r}_j-\mathbf{r}_i|/\lambda_D)}$ represents Yukawa pair potential between $i$th and $j$th particle. Here, $N$ represents the total number of particles.

\begin{figure}
  \centering
  \includegraphics[width=4.0in]{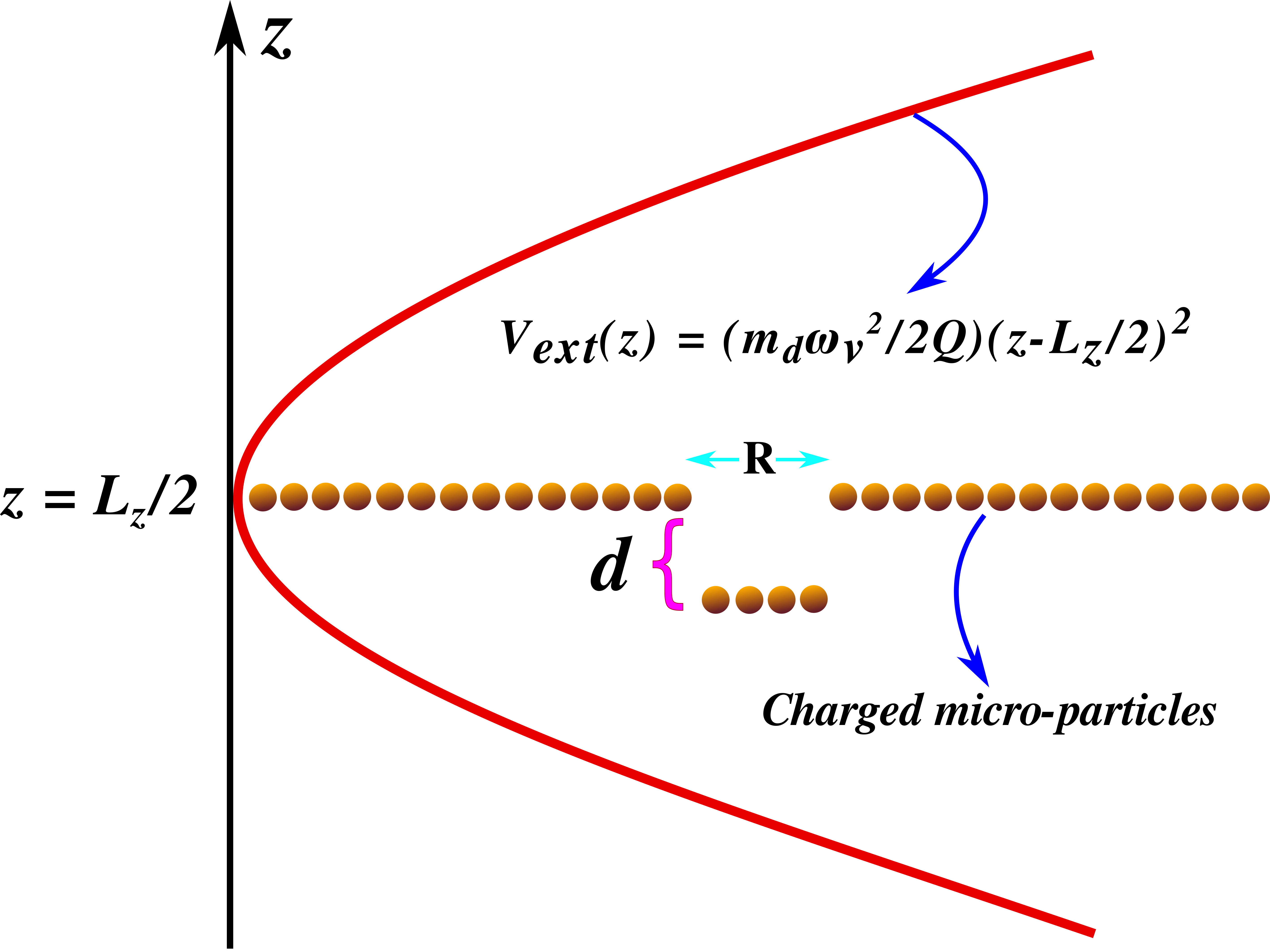}
  \caption{The schematic of the simulation setup is shown. A parabolic potential ($V_{ext}$) with a minimum at the location $z = L_z/2$ is externally applied in the vertical direction ($\hat z$). In equilibrium, charged micro-particles interacting via Yukawa pair potential form a monolayer levitating at a height $z = L_z/2$ under the effect of $V_{ext}$. Then, the particles located within a radius $R$ around the center of this monolayer are displaced vertically ($-\hat z$ direction) by a distance $d$ from their equilibrium positions. }
\label{f_sch}
\end{figure}

Initially, the system of charged particles has been relaxed to a thermal equilibrium state with a desired value of temperature $T = 300$ K. A Nose-Hoover thermostat \cite{nose1984molecular,nose1984unified,hoover1985canonical} is used for this purpose. For our chosen values of system parameters, particles in thermal equilibrium form a crystalline monolayer in the x-y plane levitating at a height $z = L_z/2$. Later, the thermostat is disconnected, and the system is allowed to evolve in a micro-canonical ensemble where the total number of particles (N), system volume (V), and total energy (E) remain conserved. Under this condition, we have imposed a perturbation to the medium, i.e., monolayer, by displacing a few particles over a distance $d$ along the $-\hat z$ direction, as illustrated by the schematic in Fig. \ref{f_sch}. It is to be noticed that only the particles which were initially located within a small circular region of radius $R$ around the center of the monolayer have been displaced.

In our study, the plasma Debye length providing the screening in the pair interactions between particles is chosen to be $\lambda_D = 5\times 10^{-3}$ m. We have considered the initial displacement to be $d = \lambda_D$ in all the cases. However, $n_d$, $\omega_v$, and $R$ have been varied in our simulation, which will be discussed in section \ref{rd}. We have chosen a particular value of $n_d = n_0 = 1\times 10^6$ m$^{-2}$ as a reference and has been used for normalization purposes. The 2D dust plasma frequency associated with $n_0$ is given by $\omega_n = \sqrt{Q^2/2\pi\epsilon_{0}m_{d}a^3}= 22.63$ $s^{-1}$. Here, $a = (\pi n_0)^{-1/2}$ represents the average inter-particle distance in the 2D monolayer. In our study, the length and time scales are normalized by $\lambda_D$ and $\omega_n^{-1}$, respectively. The velocity of the particles is normalized by the thermal velocity $v_{th} = \sqrt{k_BT/m_d}$, where $k_B$ represents the Boltzmann constant. The simulation time-step is considered to be $dt = 0.001\omega_n^{-1}$, which is small enough to track the fastest dynamics of particles. 

\section{Results and Discussion}
\label{rd}

Initially, randomly distributed charged micro-particles interacting via screened Coulomb pair potential is allowed to relax under the influence of an externally applied parabolic potential. The final equilibrium state depends on the system parameters, i.e., dust charge $Q$, dust density $n_d$, plasma Debye length $\lambda_D$, and parabolic confinement potential characterized by $\omega_v$. For our chosen values of simulation parameters, charged dust particles relax to an equilibrium state, forming a monolayer plasma crystal. This crystalline monolayer levitates inside the simulation box at a certain height of $z = L_z/2$, i.e., at the location of the minima of the external parabolic potential well, as illustrated by the schematic in Fig. \ref{f_sch}. Then, we displaced a few particles within a radius $R$ around the center of this monolayer by a distance $d$ along the vertical direction ($-\hat z$) to impose a disturbance in the medium. This has also been illustrated by the schematic in Fig. \ref{f_sch}. We have analyzed various features of our observation as a consequence of this external perturbation in the medium and presented them in the following subsections.


\subsection{Amplitude modulation via parametric decay process}\label{ampd}

The initially displaced particles exhibit vertical oscillatory motion around the x-y plane of the monolayer under the influence of the parabolic potential well. If the particles do not interact with each other, they will oscillate with a certain frequency determined by the restoring force associated with the external parabolic potential, i.e., $\omega_v$. In that case, the amplitude of their oscillatory motions will also remain constant, determined by the initial displacement $d$. However, the dynamics of each particle are strongly coupled with each other via screened Coulomb pair potential. The pair interaction strength is determined by particle density ($n_d$) and plasma Debye length ($\lambda_D$). It has been observed that the amplitude of vertical oscillations does not remain constant but gets modified periodically with time. Thus, instead of oscillating with a constant frequency and amplitude, a train of pulses (beats) with different frequencies appears in the oscillatory motions of the particles. This is a clear signature of amplitude modulation of the initially induced vertical oscillatory motion. It is worth mentioning that when we displace the entire crystalline plane vertically from its equilibrium position, all the particles oscillate with constant frequency $\omega_v$ and amplitude $d$. Thus, the amplitude modulation observed in our study occurs due to the finite boundary ($R$) of the initial perturbation. To analyze these co-related dynamics in more detail, we chose a single particle initially located at the center of the monolayer and tracked its dynamics with time. In the following subsections, we have depicted the time history of the dynamics of this particle for different cases with the changing values of system parameters, e.g., confinement frequency ($\omega_v$), dust density ($n_d$), and the radius of the initially perturbed region ($R$). 


\subsubsection{{Dependence on confinement potential}}

\begin{figure}
  \centering
  \includegraphics[width=5.0in]{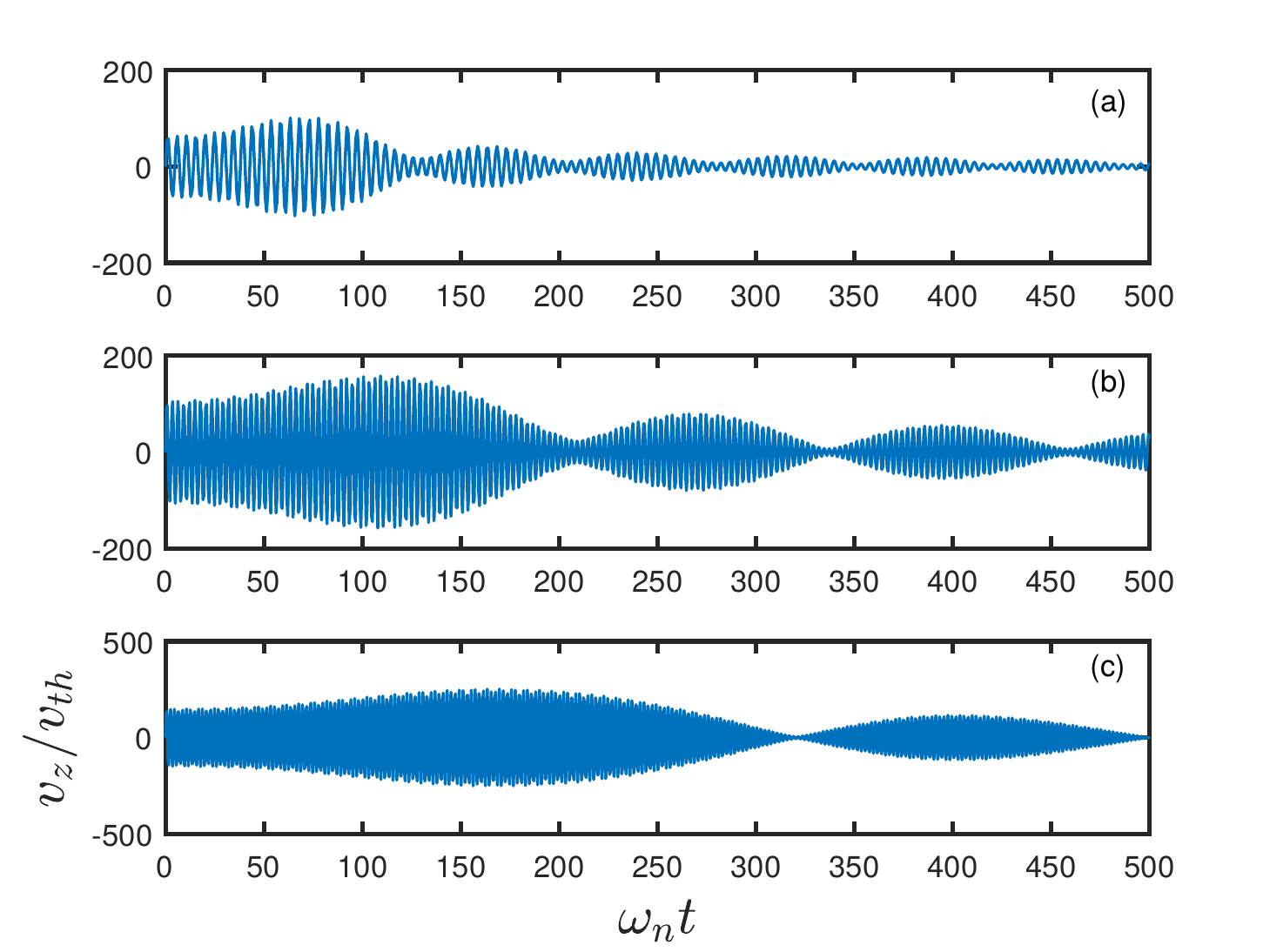}
  \caption{ Time evolution of $\hat z$-component of velocity $v_z$ of a particle, initially located at the center of the monolayer and displaced vertically, is shown for three different cases with the changing values of $\omega_v$. Here, for these three cases, the values of $\omega_v$ are chosen to be (a) $\omega_v = 1.3\omega_{n}$, (b) $\omega_v = 2.2\omega_{n}$, and (c) $\omega_v = 3.1\omega_{n}$. }
\label{vz_omgV}
\end{figure}

We performed a series of simulations with the changing values of $\omega_v$ to investigate the effect of external confinement potential on the amplitude modulation phenomena observed in our study. As stated earlier, we tracked a perturbed particle initially located at the center of the monolayer in each case. The time evolution of $\hat z$-component of velocity ($v_z$) of the tracked particle is shown in Fig. \ref{vz_omgV} for three different simulation runs with changing values of $\omega_v$. In all three cases, $v_z$ oscillates with time, and a train of pulses or beats appears in the time profile of $v_z$. It is also seen that the amplitudes of the beats are different in different cases and decay with time. Moreover, the beat frequency and the decay rate of the beat amplitude decrease as we increase the value of $\omega_v$.  

\begin{figure}
  \centering
  \includegraphics[width=5.0in]{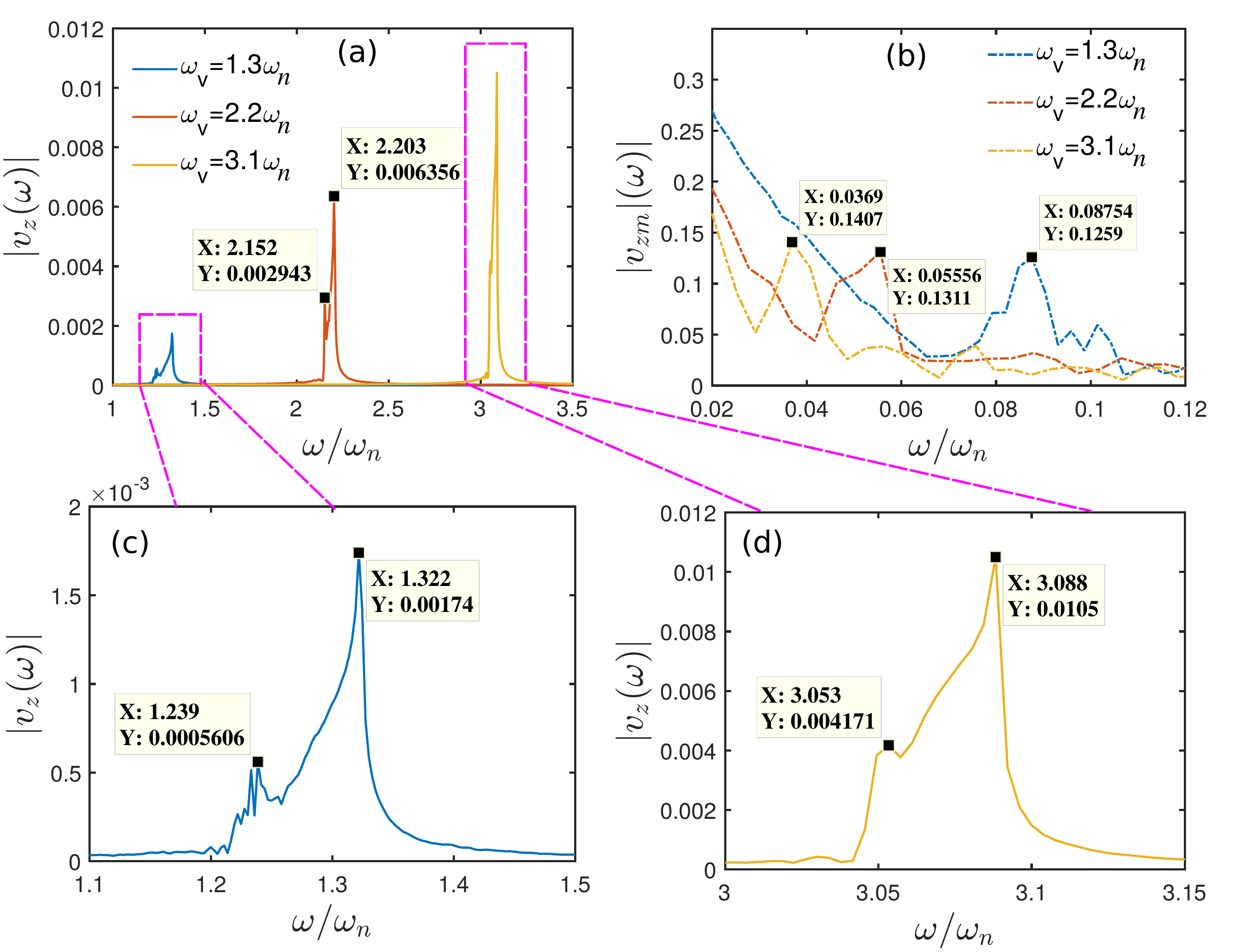}
  \caption{Fourier spectra of $v_z(t)$ are shown in subplot (a) for three different cases with $\omega_v = 1.3\omega_{n}$ (blue), $2.2\omega_{n}$ (red), and $3.1\omega_{n}$ (yellow). The corresponding Fourier spectra of $v_{zm} (t)$ (peak values of $v_z (t)$) are illustrated in subplot (b). Fourier spectra of $v_z(t)$ for $\omega_v = 1.3\omega_{n}$, $3.1\omega_{n}$ are also depicted in the zoomed scales in subplot (c) and (d), respectively.}
\label{vz_fft_omgV}
\end{figure}

To analyze the properties of these beats, we have evaluated the Fourier spectra from the time series data of $v_z$ and $v_{zm}$ of the tracked particle. Here, $v_{zm}$ represents the peak values of $v_z(t)$, and thus, the Fourier spectra of $v_{zm}(t)$ will give the information of the beat frequency. The Fourier spectra of $v_z$ and $v_{zm}$ are shown in subplots (a) and (b) of Fig. \ref{vz_fft_omgV} respectively, for three simulation runs with different values of $\omega_v$. The Fourier spectra of $v_z(t)$ for the cases of $\omega_v = 1.3\omega_{pd}$ and $3.1\omega_{pd}$ are also shown in the zoomed scales in subplots (c) and (d) of Fig. \ref{vz_fft_omgV}, respectively. From subplot (a), (c), and (d) of Fig. \ref{vz_fft_omgV}, it is clearly seen that in all three cases, instead of a single peak (at $\omega = \omega_v$), Fourier spectra of $v_z(t)$ exhibit two distinctly separated peaks appearing as a sideband of the corresponding value of $\omega_v$. However, the separation of these two peaks decreases (i.e., the sideband comes closer to the $\omega = \omega_v$) with the increase of $\omega_v$. The subplot (b) of Fig. \ref{vz_fft_omgV} illustrates that in each case, a single peak representing the beat frequency appears in the Fourier spectra of $V_{zm}(t)$. It is also seen that the beat frequency decreases with the increase of $\omega_v$ and has the same value as the difference between two peaks (sideband) appearing in the corresponding Fourier spectra of $v_z(t)$. This indicates that a parametric decay instability occurs, which is responsible for the generation of sideband in the vertical oscillatory motions of the initially perturbed particles. Consequently, a train of pulses appears as a form of amplitude modulation initiated due to the interference of these sideband frequencies. The fundamental origin of this parametric process and its dependence on $\omega_v$ will also be discussed later in this section.

\subsubsection{Dependence on dust density}

\begin{figure}
  \centering
  \includegraphics[width=5.0in]{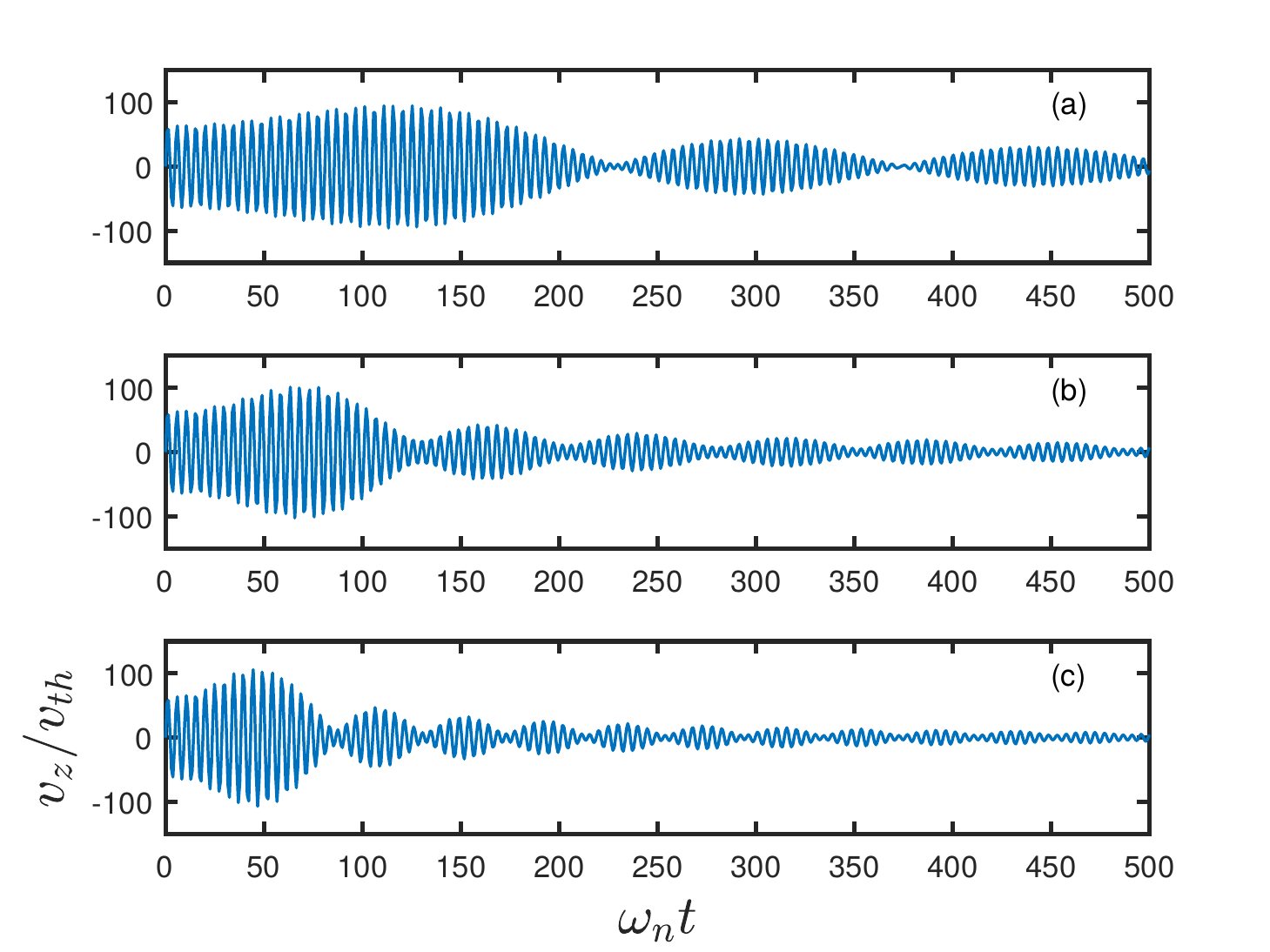}
  \caption{ The time series data of $v_z(t)$ of a particle initially located at the center of the monolayer are shown for three different cases with changing values of dust density (a) $n_d = 0.75n_0$, (b) $n_d = 1.0n_0$, and (c) $n_d = 1.25n_0$ with a fixed confinement potential frequency $\omega_v = 1.3\omega_{n}$.}
\label{vz_nd}
\end{figure}

We have also carried out simulations with changing values of dust density ($n_d$) in the 2D monolayer keeping other simulation parameters constant. In subplots (a)-(c) of Fig. \ref{vz_nd}, we have shown the time evolution of $v_z$ of the tracked particle for three different values of $n_d$. Here also, we have seen that beats appear in the time profile of $v_z$. It has been observed that the beat frequency and decay rate of the beat amplitude increase with an increase of $n_d$.

\begin{figure}
  \centering
  \includegraphics[width=5.5in]{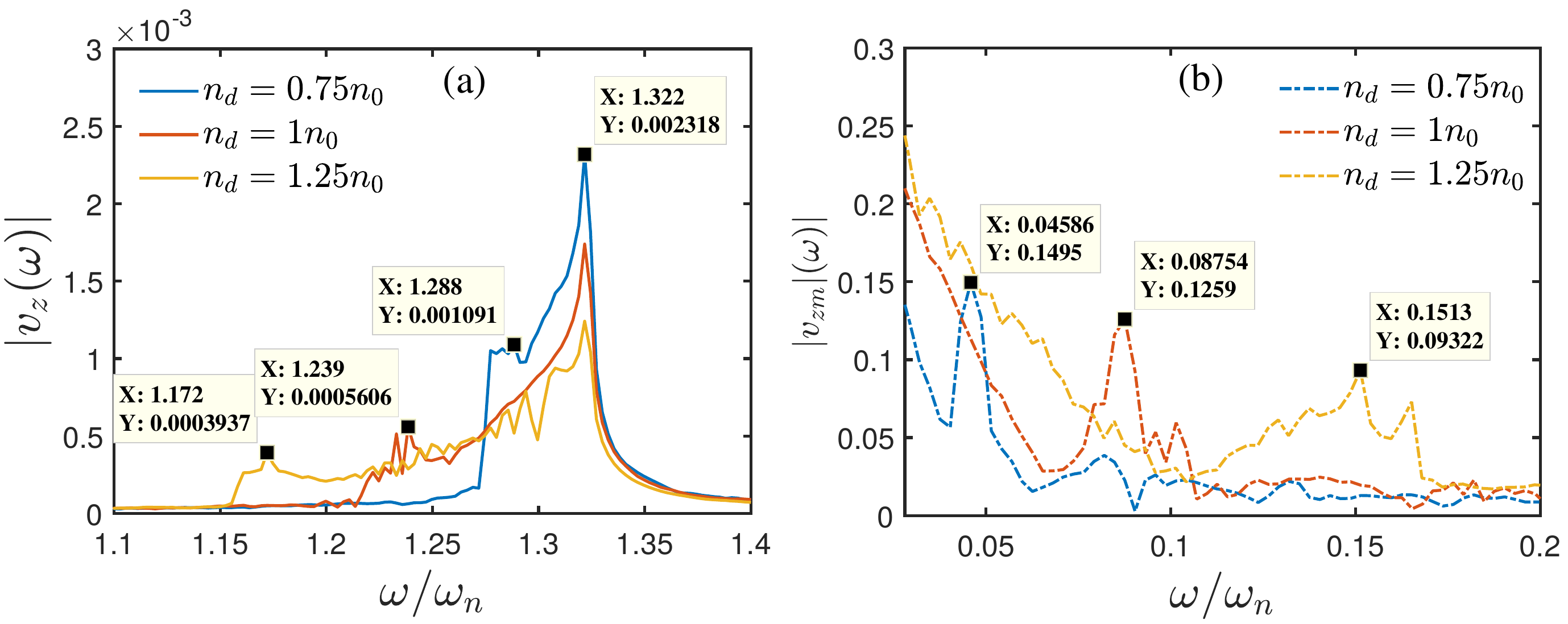}
  \caption{ Fourier spectra obtained from the time series data of $v_z(t)$ and $v_{zm}(t)$ (peak values of $v_z$) for three different cases with $n_d = 0.75n_0$ (blue), $1.0n_0$ (red), and $1.25n_0$ (yellow) are shown in subplots (a) and (b), respectively.}
\label{vz_fft_nd}
\end{figure}

To capture the parametric process involved behind the formation of the beat in the vertical oscillatory motion of particles, we have evaluated Fourier spectra of $v_z(t)$ and $v_{zm}(t)$ as before for different values of $n_d$. From the subplot (a) of Fig. \ref{vz_fft_nd}, it is seen that instead of a single peak at $\omega = \omega_v$, Fourier spectra of $v_z(t)$ in each case reveal broad spectra with two distinctly separated peaks. It is also seen that the separation between these two peaks increases with the increase of $n_d$. Fourier spectra of $v_{zm}(t)$ shown in the subplot (b) of Fig. \ref{vz_fft_nd} demonstrate that beat has a particular frequency which increases with the increase of $n_d$. Here also, it is seen that in each case, the beat frequency has the same value as the difference between two peaks that appeared in the corresponding Fourier spectra of $v_z(t)$.

\subsubsection{Dependence on perturbation radius}

\begin{figure}
  \centering
  \includegraphics[width=6.5in]{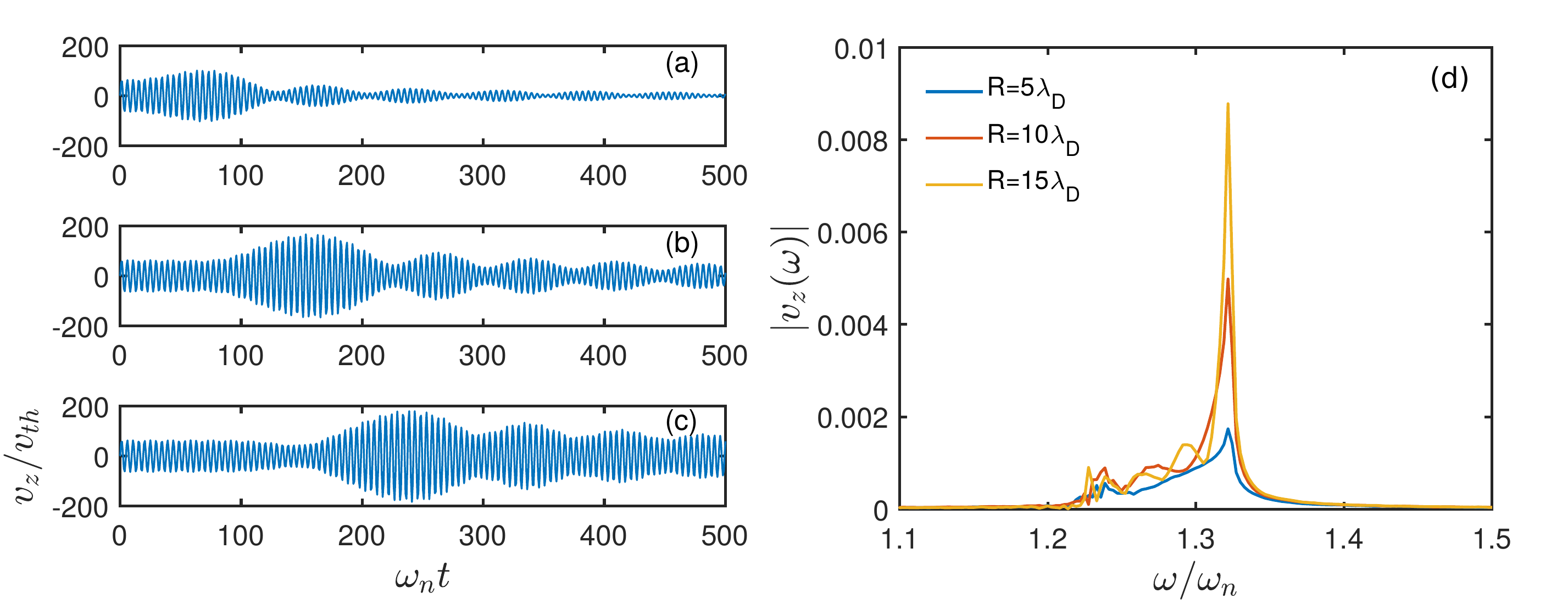}
  \caption{Time evolution of $v_z(t)$ for three different simulation runs with (a) $R = 5\lambda_D$, (b) $R = 10\lambda_D$, and (c) $R = 15\lambda_D$. In subplot (d), the corresponding Fourier spectra of $v_z(t)$ have been shown. In these simulations, the confinement frequency and particle density are kept fixed at $\omega_v = 1.3\omega_n$ and $n_d  =1.0n_0$, respectively.}
\label{vz_R}
\end{figure}

As mentioned earlier, the amplitude modulation phenomena initiate only when the initial perturbed region's size is less than that of the monolayer. When the entire monolayer is perturbed, the whole plane exhibits sinusoidal oscillation around the equilibrium height, i.e., $z = L_z/2$, without forming any beat or train of pulses in the particle's motion. However, to study the effect of the perturbation area, we have performed a few simulations with changing values of radius ($R$) of the initially perturbed region. The time evolution of $v_z(t)$ and the corresponding Fourier spectra of a tracked particle initially located at the center of the monolayer has been shown in Fig. \ref{vz_R} for three different values of $R$. It has been observed that the modulated amplitude becomes higher for higher values of $R$, as can be seen from the subplots (a)-(c) of Fig. \ref{vz_R}. It is also seen that for higher values of $R$, modulation occurs at a later time. This is a consequence of the fact that amplitude modulation initiates at the boundary of the perturbed and unperturbed regions. Thus, for higher values of $R$, it takes longer times to reach the information at the center of the monolayer where the chosen particle is located. However, the beat frequency does not change with the radius ($R$) of the perturbed region, as clearly depicted in subplot (d) of Fig. \ref{vz_R}.

The effect of initial displacement ($d$) on the properties of the medium has been reported in an earlier study where we have shown that crystalline monolayer melts through a first-order phase transition above a threshold value of $d$ \cite{maity2022parametric}. In that study, it has also been shown that this threshold value of $d$ for which the phase transition occurs depends upon the value of $R$.  

\begin{figure}
  \centering
  \includegraphics[width=4.5in]{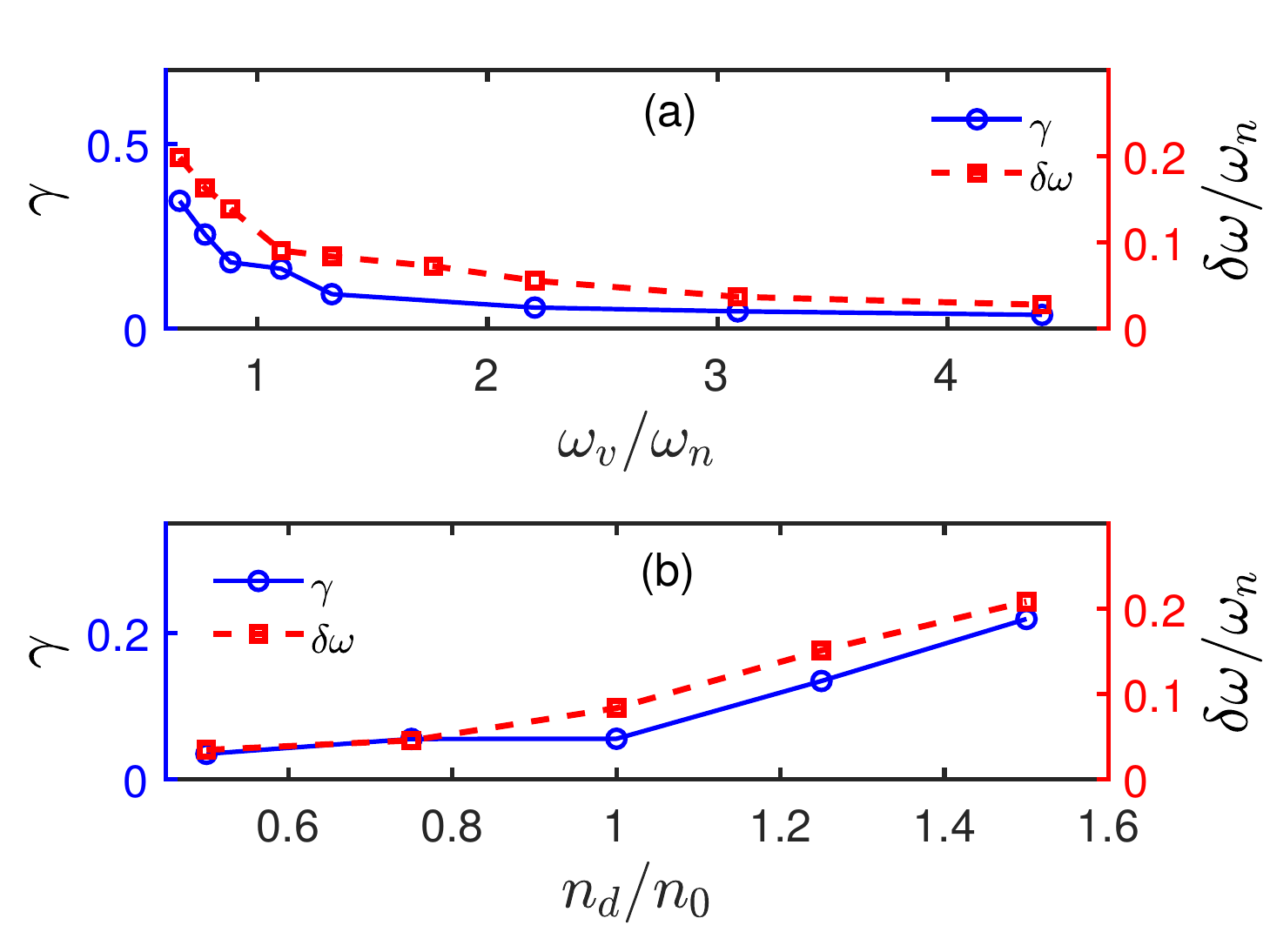}
  \caption{The variation of beat amplitude decay rate ($\gamma$) and beat frequency ($\delta \omega$) as a function of (a) confinement frequency $\omega_v$ and (b) particle density $n_d$ in the 2D monolayer .}
\label{beat_fg}
\end{figure}


Let us now summarize all the results so far that have been presented and understand the fundamental origin behind the amplitude modulation phenomena observed in our study. As we impose an external perturbation by displacing a few particles initially located in the monolayer's central regime, they oscillate vertically under parabolic confinement potential. However, the dynamics of particles are strongly coupled with each other via pair interactions. As a result, a shear stress is generated between the neighboring particles, which triggers a nonlinear parametric process. Consequently, sidebands develop (Fig. \ref{vz_fft_omgV} $\&$ \ref{vz_fft_nd}) in the frequency spectra of vertical oscillatory motions of the particles. Therefore, amplitude modulation occurs, and beats generate through the interference of these sidebands. It is important to note that beat is formed not only in the dynamics of perturbed particles but also in the initially undisturbed region of the crystalline plane, as will be discussed in section \ref{swg}. The beat amplitude erodes in time as the kinetic energy of the perturbed particles is transported and distributed in the whole crystalline plane. With the increase of parabolic confinement frequency ($\omega_v$), the amplitude of the oscillating velocity (i.e., kinetic energy) of perturbed particles increases. This causes the effective interaction between the neighboring particles to be less efficient and reduces the shear stress. As a result, beat frequency and beat amplitude decay rate decrease with the increase of $\omega_v$. This has been depicted in subplot (a) of Fig. \ref{beat_fg} and also can be seen from Fig. \ref{vz_omgV} $\&$ \ref{vz_fft_omgV}. However, the average inter-particle separation decreases as we increase the particle density ($n_d$) in the 2D monolayer. As a result, effective pair interaction strength between neighboring particles increases. Consequently, the nonlinearity in the parametric process responsible for the amplitude modulation increases. Therefore, beat frequency and beat amplitude decay rate increases with $n_d$. This is clearly illustrated in subplot (b) of Fig. \ref{beat_fg} and also shown in Fig. \ref{vz_nd} $\&$ \ref{vz_fft_nd}. 

\subsection{Theoretical model}
\label{model}

\begin{figure}
  \centering
  \includegraphics[width=5.5in]{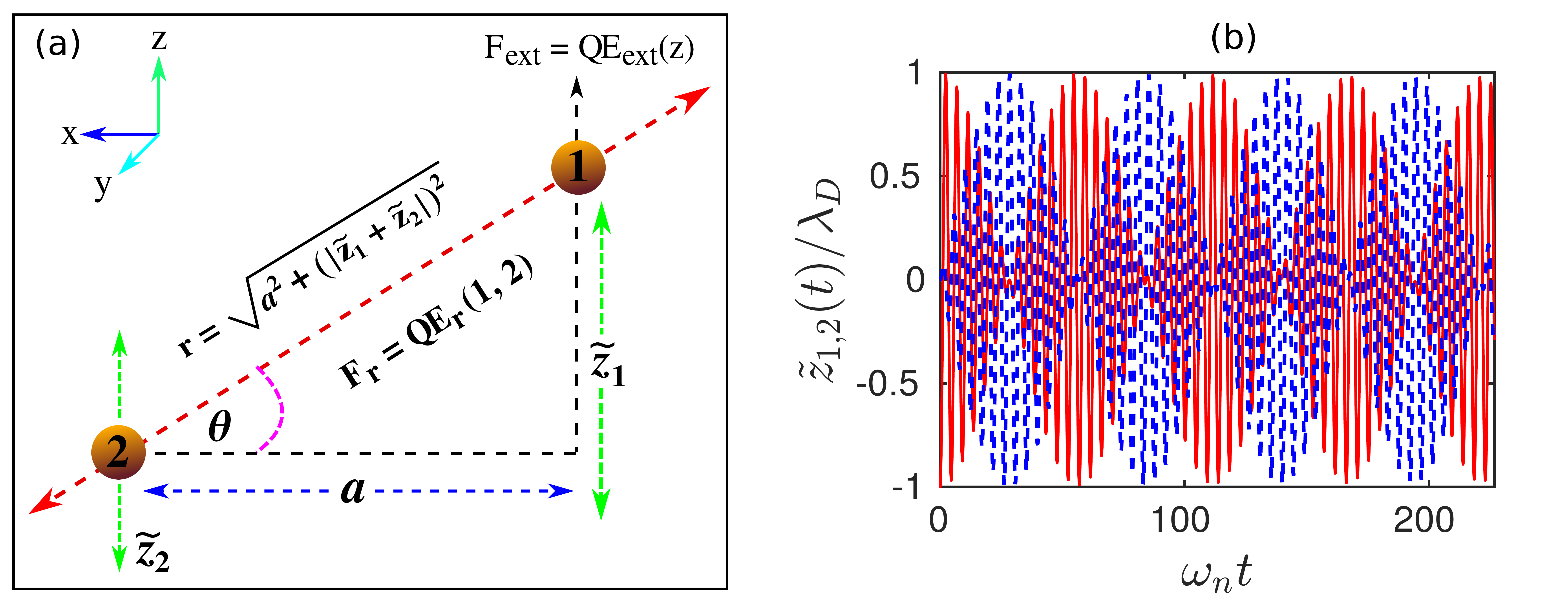}
  \caption{ A schematic of the theoretical model has been shown in subplot (a). Time evolution of $z$-coordinates of particle 1 ($\tilde{z}_1$) and particle 2 ($\tilde{z}_2$) obtained from the theoretical model have been shown in subplot (b).}
\label{model_schmtic}
\end{figure}

\begin{figure}
  \centering
  \includegraphics[width=5.5in]{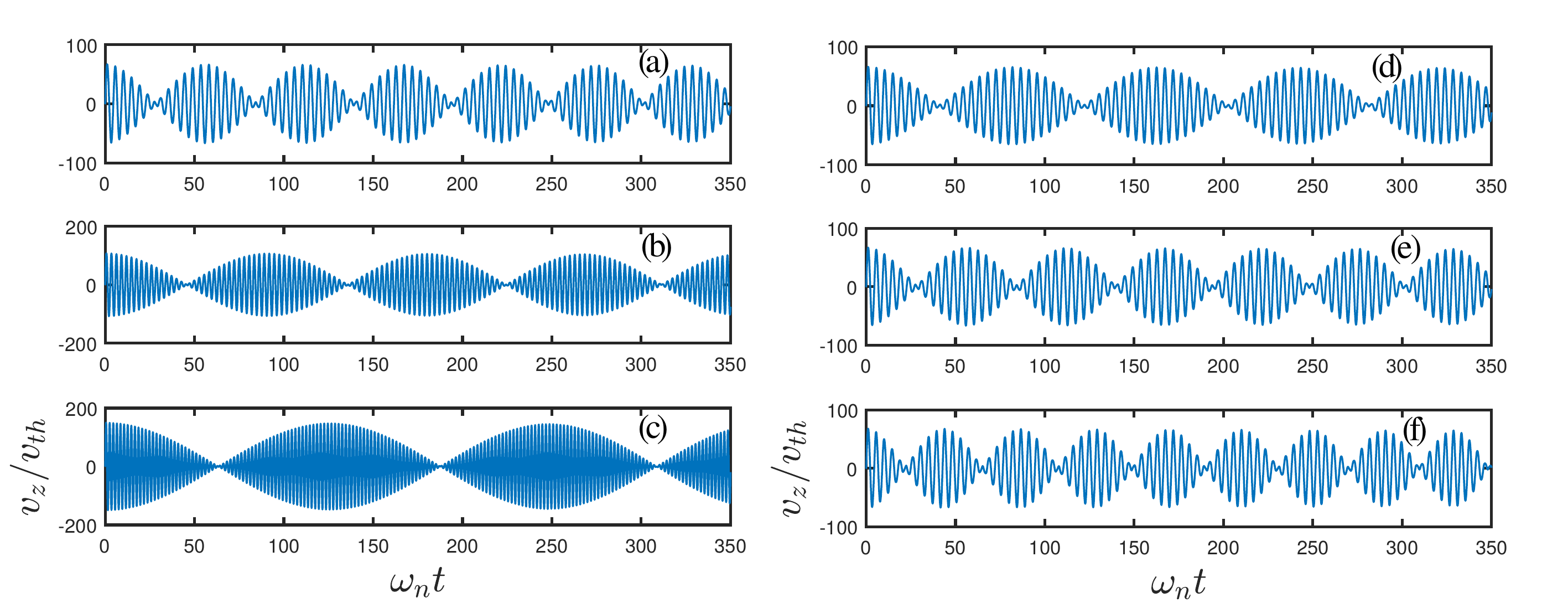}
  \caption{ Time evolution of $v_z(t)$ obtained from the theoretical model for three different values of confinement frequency, (a) $\omega_v = 1.3\omega_{n}$, (b) $2.2\omega_{n}$, and (c) $3.1\omega_{n}$ with a fixed $n_d = 1.0n_0$ have been shown. The same has been illustrated for three different values of dust density in subplots (d) $n_d = 0.75n_0$, (e) $n_d = 1.0n_0$, and (f) $n_d = 1.25n_0$ with a fixed $\omega_v = 1.3\omega_{n}$. }
\label{model_subplot}
\end{figure}

We have also developed a simple theoretical model based on the arguments given above, which support our observations qualitatively. In our model, we have considered two particles coupled with each other via Yukawa pair interaction in the presence of an external parabolic potential ($V_{ext}(z)$). One particle (particle 1) is subjected to an initial displacement in the vertical direction by a distance $z_1(t = 0) = -\lambda_D$ as in the case of MD simulations. The other particle (particle 2) is kept unperturbed initially. Thus, our model approximately mimics the boundary between perturbed and unperturbed regions of the monolayer. The schematic of the model is depicted in subplot (a) of Fig. \ref{model_schmtic}. The equations of motion of these two particles along the vertical ($\hat z$) direction can be expressed as,

\begin{equation}
m_d\frac{d^2\tilde{z}_1}{dt^2} = -m_d\omega_v^2\tilde{z}_1 - \frac{Q^2}{4\pi\epsilon_0}\left[\frac{1}{r^2} + \frac{1}{r\lambda_D}\right]\exp{\left(-r/\lambda_D\right)}\sin\theta
\label{eq1}
\end{equation}

\begin{equation}
m_d\frac{d^2\tilde{z}_2}{dt^2} = -m_d\omega_v^2\tilde{z}_2 - \frac{Q^2}{4\pi\epsilon_0}\left[\frac{1}{r^2} + \frac{1}{r\lambda_D}\right]\exp{\left(-r/\lambda_D\right)}\sin\theta
\label{eq2}
\end{equation}

, where $\sin\theta = \left(\tilde{z}_1+\tilde{z}_2\right)/r$ with $r = \sqrt{a^2 + (|\tilde{z}_1^2 + \tilde{z}_2^2|)}$ representing the radial distance between the two particles. The first terms of the right-hand side (RHS) of equations \ref{eq1} and \ref{eq2} represent the force $\mathbf{F}_{ext}$ associated with the external parabolic potential, $V_{ext}(z) = (m_d\omega_v^2/2Q)z^2$. The second terms of the RHS of these two equations represent the $\hat z$-component of the force $\mathbf{F}_r$ associated with Yukawa pair interaction between particles. We have solved equation \ref{eq1} and equation \ref{eq2} numerically and shown the time evolution of $\tilde{z}_1$ and $\tilde{z}_2$ in subplot (b) of Fig. \ref{model_schmtic}. As in the case of MD simulations, it is seen that beat is formed in the time evolution of both perturbed and unperturbed particles. We also did a parametric study using our theoretical model, and the results are shown in Fig. \ref{model_subplot}. It is seen that beat frequency decreases with the increase of $\omega_v$, as shown in the subplots (a)-(c) of Fig. \ref{model_subplot}. Whereas, beat frequency increases with an increase of $n_d$, as have been shown in subplots (d)-(f) of Fig. \ref{model_subplot}. Here, the dependence of $n_d$ is revealed through the parameter $a = \sqrt{1/\pi n_d}$ representing the average inter-particle distance in the 2D crystalline plane. Thus, all the simulation findings qualitatively agree with the numerical results obtained from our theoretical model.


\subsection{Surface wave generation}
\label{swg}

\begin{figure}
  \centering
  \includegraphics[width=5.5in]{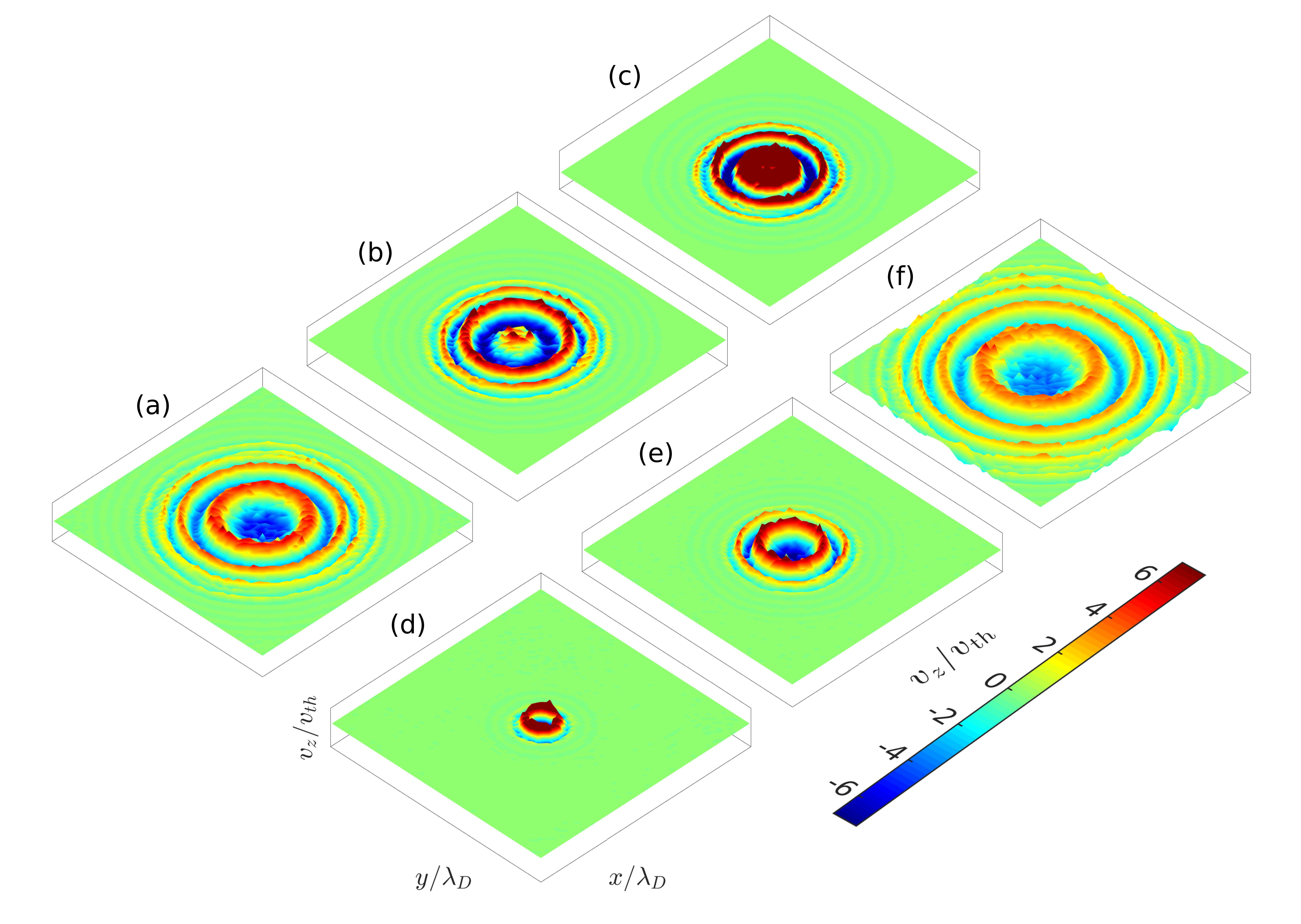}
  \caption{Distributions of $v_z$ in the $xy$-plane of the monolayer is shown at a particular instant of time $\omega_nt = 900$ for different cases with the changing values of $\omega_v$ and $n_d$. In subplots (a)-(c) the values of $\omega_v$ are chosen to be $\omega_v = 1.3\omega_n$, $1.8\omega_n$, and $2.2\omega_n$, respectively for a fixed value value of $n_d = 1.0n_0$. In subplots (d)-(e), $\omega_v$ is kept constant at $\omega_v = 1.3\omega_n$ with the changing values of dust density $n_d = 0.5n_0$, $0.75n_0$, and $1.15n_0$, respectively.}
\label{vz_space}
\end{figure}

\begin{figure}
  \centering
  \includegraphics[width=5.5in]{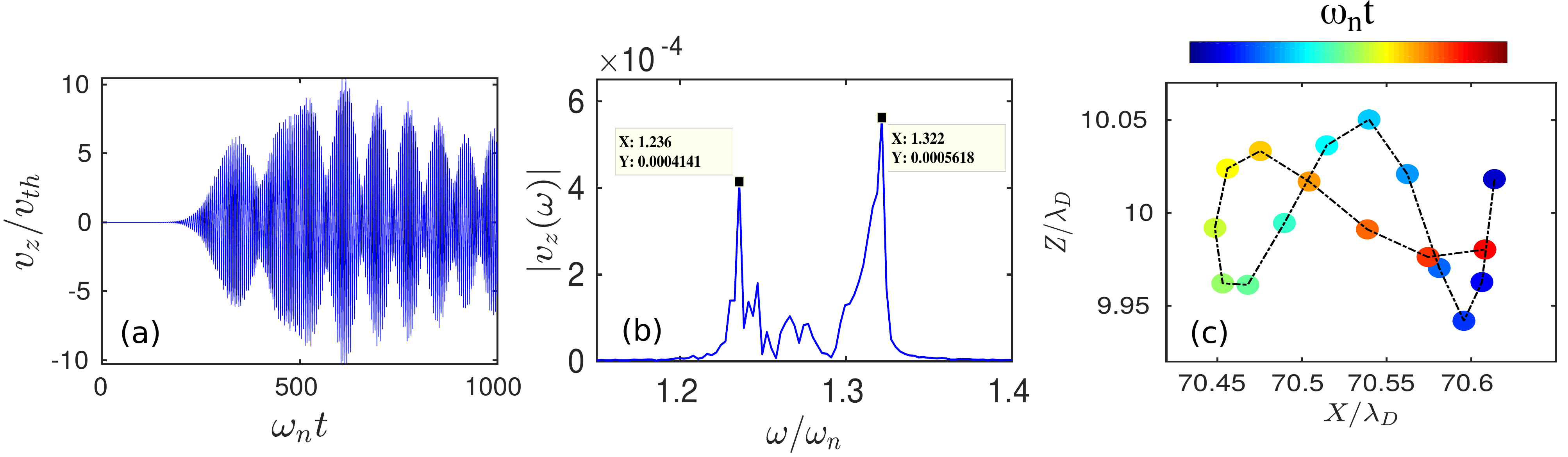}
  \caption{ Time evolution of $v_z$ of a tracked particle located far away from the initially perturbed region ($R = 5\lambda_D$), i.e., $(x, y) = ((x_c - 30\lambda_D$), $y_c$) is shown in subplot (a). Here, ($x_c$, $y_c$) represents the x-y coordinate of the center of the monolayer. In subplot (b), the Fourier spectra of $v_z(t)$ is depicted. The trajectory of the tracked particle in the $x-z$ plane in a chosen period of time is illustrated in subplot (c). Here, the color symbols from blue to red represent the evolution of time.}
\label{far_particle}
\end{figure}

\begin{figure}
  \centering
  \includegraphics[width=4.5in]{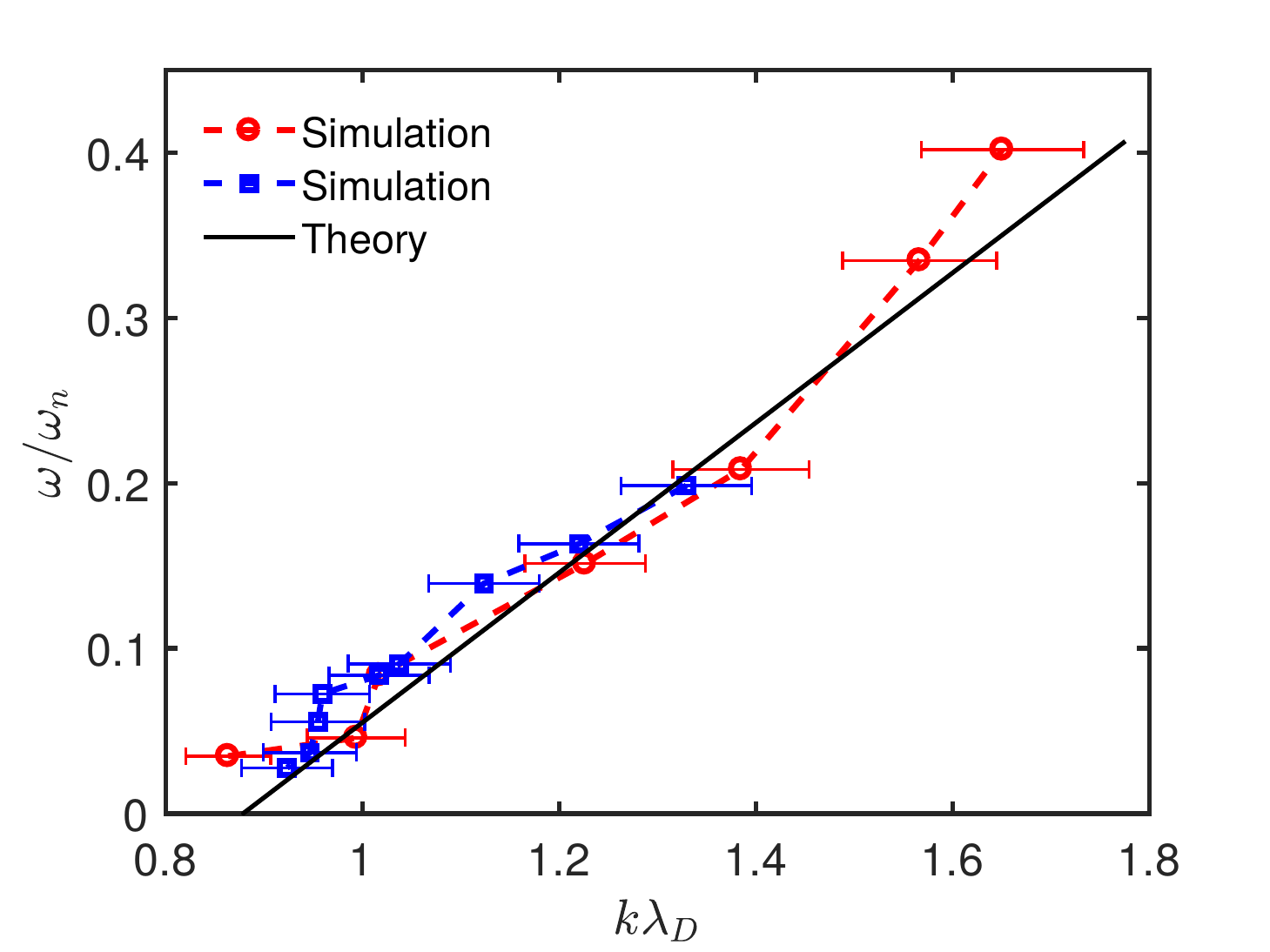}
  \caption{Dispersion relation of the transverse surface wave in $\omega-k$ plane. The dispersion properties obtained by changing the values of $n_d$ with a fixed $\omega_v = 1.3\omega_n$ are shown by the red circle-marked dashed line. The dispersion relation illustrated by a blue square-marked dashed line is for the case where we change the values of $\omega_v$ with a fixed dust density $n_d = 1.0n_0$.}
\label{disprs}
\end{figure}

It would be interesting to look at how the rest of the monolayer, which was initially unperturbed, responds to the external disturbance imposed in the medium. The distributions of $v_z$ in the x-y plane of the monolayer have been shown in Fig. \ref{vz_space} for different values of $\omega_v$ and $n_d$ at a particular instant of time $\omega_nt = 900$. It is seen that in all the cases, the medium responds collectively to the initial external perturbation. It is seen that concentric circular wavefronts are generated in the profile of $v_z$ around the initially perturbed region and spread over in the x-y plane of the crystalline layer. Moreover, at a particular instant of time, the area over which these circular wavefronts have been spread in the monolayer depends upon the values of $\omega_v$ and $n_d$, as can be seen from the subplots (a)-(f) of Fig. \ref{vz_space}.
This is the consequence of the fact that the surface wave is generated through the formation of beat and beat frequency changes for different values $\omega_v$ and $n_d$.

It would also be interesting to see the particle-level dynamics while these circular waves propagate through the crystalline plane of the medium. For this purpose, we choose to track a particle initially located at a radial distance $r = 30\lambda_D$ from the center of the monolayer. It is to be noted that in this case, the radius of the initially perturbed region is to be $R = 5\lambda_D$. Thus, our chosen particle is located far away from the initial perturbation region. The dynamics of this tracked particle have been illustrated in Fig. \ref{far_particle}. Time evolution of $v_z$ of the tracked particle reveals the formation of beat, as depicted in subplot (a) of Fig. \ref{far_particle}. Thus, the beat occurs not only in the dynamics of initially perturbed particles but also for particles located in the initially undisturbed region of the monolayer. This has also been predicted from our theoretical model. It is also seen from the subplot (a) of Fig. \ref{far_particle} that the beat appears only after a certain time, which is the time the first wavefront takes to reach the location of the tracked particle. The Fourier spectra of $v_z(t)$, shown in subplot (b) of Fig. \ref{far_particle}, reveal that the vertical motion of the tracked particle is associated with two different frequencies. It is also seen that the difference between these two frequencies representing the beat frequency is the same as the perturbed particles, as shown in the subplot (b) (blue dashed line) of Fig. \ref{vz_fft_omgV}. Thus, the circular wavefronts propagating through the surface of the monolayer are nothing but the beat waves that originated due to the amplitude modulation of initial perturbation. The dynamics of the tracked particle in the x-z plane are demonstrated in subplot (c) of Fig. \ref{far_particle} over the duration of a beat period. The position of the particle in the x-z plane at different times has been represented by the blue to red color symbols. It is seen that the particle exhibits oscillatory motion in the x-z plane. However, the mean position of the particle over a beat period almost remains unchanged. This reveals that the particle's motion has both transverse and longitudinal components, which is a typical characteristic of a surface wave. 

The dispersion property of the surface wave observed in our study has been depicted in Fig. \ref{disprs} in the $\omega-k$ plane. The dispersion relation has been obtained by measuring the wavelength and frequency (beat frequency, $\delta \omega$) of the 
 fully developed surface waves with changing values of system parameters, i.e., $\omega_v$ and $n_d$. In both cases, we have obtained a linear dispersion relation between frequency $\omega$ and wavenumber $k$, as shown by the blue squares marked line and red circles marked line in Fig. \ref{disprs}, respectively. It is also seen that the dispersion relation obtained from our simulations closely matches the theoretical dispersion curve of the transverse shear wave shown by the solid black line in Fig. \ref{disprs}. The theoretical dispersion relation of the transverse shear wave has been obtained from the relation, $\omega \approx c_sk$, where $c_s = \sqrt{(k_BT/m_d)\Gamma\exp{-(a/2\lambda_D)}}$ is the velocity of the shear wave. Here, $\Gamma = Q^2/4\pi\epsilon_0ak_BT$ represents the Coulomb coupling parameter. The measured group velocity of the surface wave observed in our simulation is $c_{sim}\sim5.1~mm/s$, which is close to the theoretically estimated shear wave velocity,  $c_{s}\sim5.3~mm/s$. Thus, the surface waves observed in our study are the out-of-plane transverse shear waves generated due to the velocity shear stress between neighboring particles.


\section{Summary}
\label{summary}

In this study, we have investigated the response of a monolayer plasma crystal to an external perturbation under various conditions. We have performed 3D molecular dynamics simulations of a system of charged micro-particles (dust) interacting via Yukawa pair interactions. In our simulations, we have also considered an external parabolic potential along the vertical ($\hat z$) direction which mimics the combined effect of gravity and sheath electric field typically present in a dusty plasma experiment. It has been shown that a 2D monolayer of charged dust particles can be formed for a suitable choice of parabolic confinement frequency ($\omega_v$) and dust density ($n_d$). We then imposed an external perturbation in the medium by displacing a few particles within a small circular region around the center of the monolayer along the vertical ($\hat z$) direction. We have analyzed in detail the response of the medium to this externally imposed perturbation under various conditions, e.g., changing values of $\omega_v$, $n_d$, and $R$. It has been shown that the induced vertical oscillatory motion of the perturbed particles gets modulated through a parametric decay process initiated due to the shear stress between neighboring particles. Consequently, beat generates in the dynamics of both perturbed and initially unperturbed particles. As a result, concentric circular wavefronts are created, which propagate through the surface of the monolayer in the radially outward direction from the initially perturbed region. We have shown that these surface waves follow the dispersion relation of a transverse shear wave. A simple theoretical model has been provided supporting our simulation observations. An experimental realization of our simulation findings would be interesting. 
 
%

\section*{References}
\bibliographystyle{unsrt}
\bibliography{sw.bib}
\end{document}